\documentclass[aps,nature,twocolumn,superscriptaddress,nopacs,amsmath,amssymb,letter,citeautoscript]{revtex4}%

\setcitestyle{super}
\usepackage{color}
\usepackage{graphicx}% Include figure files
\usepackage{dcolumn}% Align table columns on decimal point
\usepackage{bm}% bold math

\begin{document}

\title{Direct observation of spin-polarised bulk bands in\\ an inversion-symmetric semiconductor}
\author{J.~M.~Riley}
\affiliation {SUPA, School of Physics and Astronomy, University of
St. Andrews, St. Andrews, Fife KY16 9SS, United Kingdom}

\author{F.~Mazzola}
\affiliation{Department of Physics, Norwegian University of Science and Technology (NTNU), N-7491 Trondheim, Norway}

\author{M.~Dendzik}
\author{M.~Michiardi}
\affiliation{Department of Physics and Astronomy, Interdisciplinary Nanoscience Center (iNANO), Aarhus University, 8000 Aarhus C, Denmark}

\author{T.~Takayama}
\affiliation{Department of Physics, University of Tokyo, Hongo, Tokyo 113-0033}
\affiliation{Max Planck Institute for Solid State Research, 70569 Stuttgart, Germany}

\author{L.~Bawden}
\affiliation {SUPA, School of Physics and Astronomy, University of
St. Andrews, St. Andrews, Fife KY16 9SS, United Kingdom}

\author{C.~Graner{\o}d}
\affiliation{Department of Physics, Norwegian University of Science and Technology (NTNU), N-7491 Trondheim, Norway}

\author{ M.~Leandersson}
\author{T.~Balasubramanian}
\affiliation{MAX IV Laboratory, Lund University, P. O. Box 118, 221 00 Lund, Sweden}

\author{M.~Hoesch}
\author{T.~K.~Kim}
\affiliation{Diamond Light Source, Harwell Campus, Didcot, OX11 0DE, United Kingdom}

\author{H. Takagi}
\affiliation{Department of Physics, University of Tokyo, Hongo, Tokyo 113-0033}
\affiliation{Max Planck Institute for Solid State Research, 70569 Stuttgart, Germany}

\author{W. Meevasana}
\affiliation {School of Physics, Suranaree University of Technology, Nakhon Ratchasima, 30000, Thailand} 
\affiliation{NANOTEC-SUT Center of Excellence on Advanced Functional Nanomaterials, Suranaree University of Technology, Nakhon Ratchasima 30000, Thailand}

\author{Ph.~Hofmann}
\affiliation{Department of Physics and Astronomy, Interdisciplinary Nanoscience Center (iNANO), Aarhus University, 8000 Aarhus C, Denmark}

\author{M.~S.~Bahramy}
\affiliation{Quantum-Phase Electronics Center and Department of Applied Physics, The University of Tokyo, Tokyo 113-8656, Japan}
\affiliation{RIKEN center for Emergent Matter Science (CEMS), Wako 351-0198, Japan}

\author{J.~W.~Wells}
\affiliation{Department of Physics, Norwegian University of Science and Technology (NTNU), N-7491 Trondheim, Norway}

\author{P.~D.~C. King}
\altaffiliation {To whom correspondence should be addressed: philip.king@st-andrews.ac.uk}
\affiliation {SUPA, School of Physics and Astronomy, University of
St. Andrews, St. Andrews, Fife KY16 9SS, United Kingdom}

\date{\today}% It is always \today, today,
             %  but any date may be explicitly specified

\maketitle 

{\bf Methods to generate spin-polarised electronic states in non-magnetic solids are strongly desired to enable all-electrical manipulation of electron spins for new quantum devices.~\cite{koo_control_2009} This is generally accepted to require breaking global structural inversion symmetry.~\cite{koo_control_2009,hsieh_topological_2008,Bahramy:NatCommun:3(2012)1159--,Murakawa_Detection_2013,mourik_signatures_2012} In contrast, here we present direct evidence from spin- and angle-resolved photoemission spectroscopy for a strong spin polarisation of bulk states in the centrosymmetric transition-metal dichalcogenide WSe$_2$. We show how this arises due to a lack of inversion symmetry in constituent structural units of the bulk crystal where the electronic states are localised, leading to enormous spin splittings up to $\sim\!0.5$~eV, with a spin texture that is strongly modulated in both real and momentum space. As well as providing the first experimental evidence for a recently-predicted `hidden' spin polarisation in inversion-symmetric materials,~\cite{zhang_hidden_2014} our study sheds new light on a putative spin-valley coupling in transition-metal dichalcogenides,~\cite{mak_control_2012,zeng_valley_2012,zhang_electrically_2014} of key importance for using these compounds in proposed valleytronic devices.}

The powerful combination of inversion symmetry [$E(\mathbf{k},\uparrow)=E(-\mathbf{k},\uparrow)$] with time-reversal symmetry  [$E(\mathbf{k},\uparrow)=E(-\mathbf{k},\downarrow)$] ensures that electronic states of non-magnetic centrosymmetric materials must be doubly spin-degenerate. If inversion symmetry is broken, however, relativistic spin-orbit interactions can induce a momentum-dependent spin splitting via an effective magnetic field imposed by spatially-varying potentials. If the resulting spin polarisations can be controllably created and manipulated, they hold enormous promise to enable a range of new quantum technologies. These include routes towards electrical control of spin precession for spin-based electronics,~\cite{datta_electronic_1990,koo_control_2009} new ways to engineer topological states~\cite{das_engineering_2013,zhou_engineering_2014} and possible hosts of Majorana fermions for use in quantum computation.~\cite{mourik_signatures_2012} To date, there are two generally-accepted categories of materials in which spin-polarised states can be stabilised without magnetism. The first exploits the breaking of structural inversion symmetry of a centrosymmetric host by imposing an electrostatic potential gradient, for example within an asymmetric quantum well, leading to Rashba-split~\cite{bychkov_properties_1984} states localised at surfaces or interfaces.~\cite{nitta_gate_1997,ast_giant_2007,king_large_2011,king_STO} In the second, a lack of global inversion symmetry in the unit cell mediates spin splitting of the bulk electronic states, either through a Dresselhaus-type interaction,~\cite{dresselhaus_spin-orbit_1955} or a recently discovered bulk form of the Rashba effect.~\cite{ishizaka_giant_2011,Murakawa_Detection_2013} 

\begin{figure*}[t]
\begin{center}
\includegraphics[width=\textwidth]{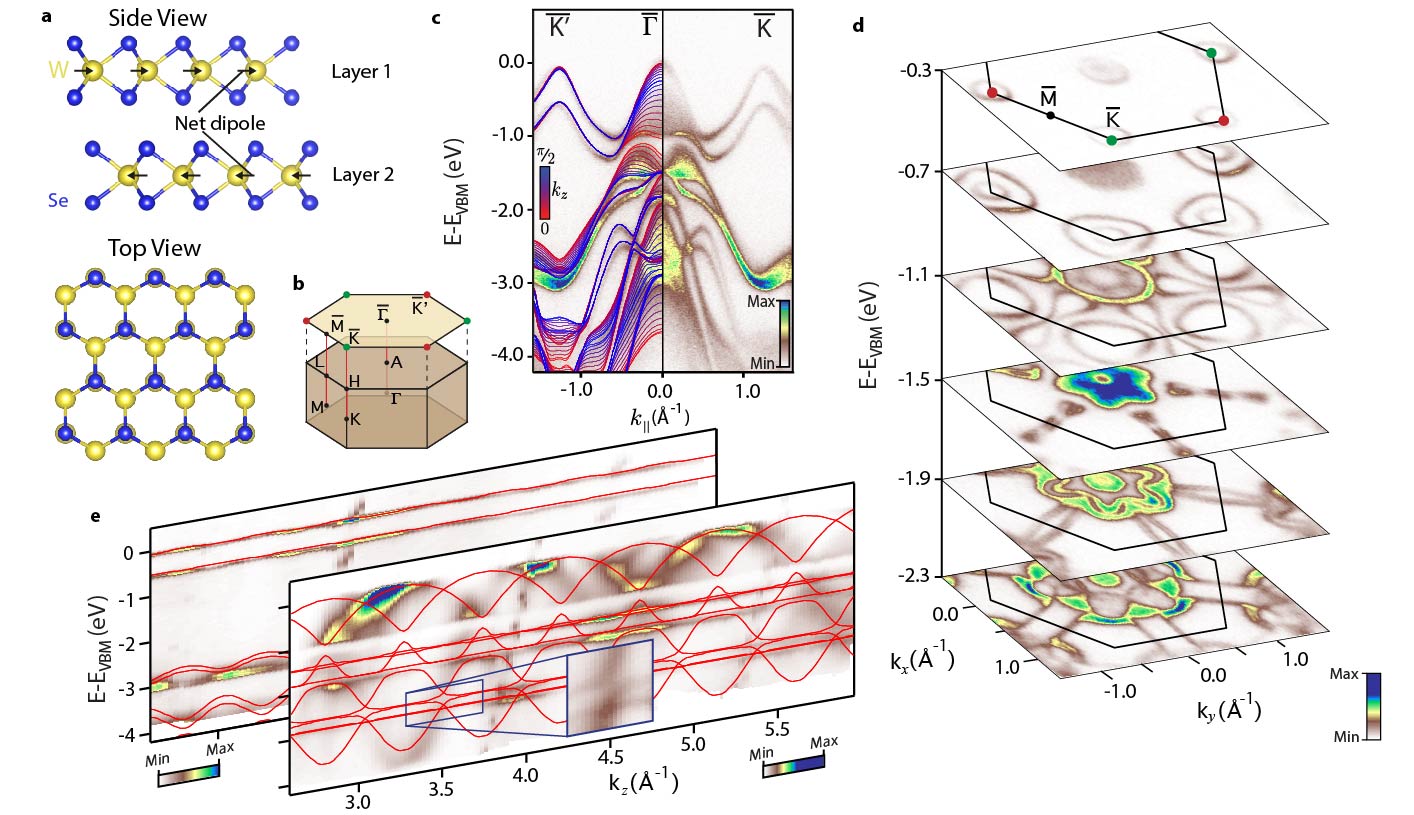}
\caption{ \label{f:overview} {\bf Bulk electronic structure of WSe$_2$.} (a) Side and top views of the bulk crystal structure of WSe$_2$. The unit cell contains two Se-W-Se units in which there is a net in-plane dipole pointing to the right and left, respectively. (b) Corresponding bulk and surface Brillouin zone. ARPES measurements ($h\nu=125$~eV) of (c) the electronic structure along the $\overline{K'}-\overline{\Gamma}-\overline{K}$ direction and (d) isoenergy contours throughout the surface Brillouin zone, reveal sharply-defined bands (for example the upper valence bands at $\overline{K}$) with significant in-plane dispersion, indicative of two-dimensional electronic states. We also observe broader ``filled-in'' pockets of spectral weight characteristic of three-dimensional states, where the finite $k_z$ resolution of ARPES leads to broadening. (e) We directly confirm this absence or presence, respectively, of significant $k_z$ dispersion from photon energy-dependent ARPES measurements (see also methods and Supplementary Fig.~S1). Our measured electronic structure is in excellent agreement with that calculated from density-functional theory (solid lines in (c) and (e)), confirming that we are probing the bulk electronic state of WSe$_2$.}
\end{center}
\end{figure*}

Here, we present the first experimental observation of a third distinct class: a material which has bulk inversion symmetry but nonetheless exhibits a large spin polarisation of its bulk electronic states. We demonstrate this for the transition-metal dichalcogenide $2H$-WSe$_2$. This layered compound is composed of stacked Se-W-Se planes  (Fig.~\ref{f:overview}(a)), each of which contains an in-plane net dipole moment which is proposed to lead to a strong spin-valley coupling for an isolated monolayer.~\cite{xiao_coupled_2012,mak_control_2012,zeng_valley_2012}  The bulk unit cell contains two such monolayers, stacked in a staggered `AB' configuration, restoring inversion symmetry and necessitating spin degeneracy of the bulk electronic states. Nevertheless, combining spin- and angle-resolved photoemission spectroscopy (ARPES) with electronic structure calculations, we uncover a large layer- and momentum-dependent spin polarisation of these bulk bands. This first experimental observation of spin-polarised bulk bands in an inversion-symmetric non-magnetic material opens unique possibilities to selectively tune spin populations in a wide variety of new materials.
\begin{figure*}[t]
\begin{center}
\includegraphics[width=\textwidth]{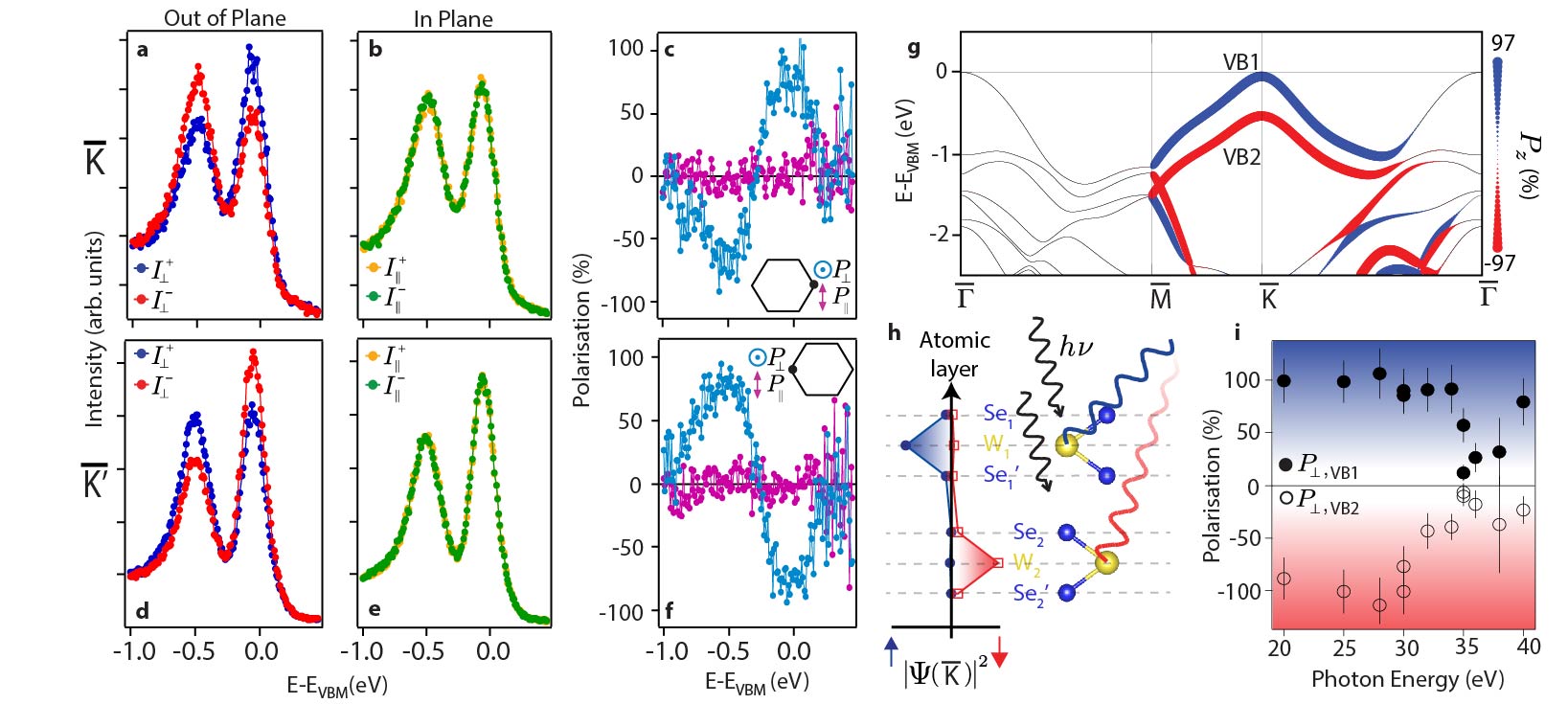}
\caption{ \label{f:Kpoint} {\bf Observation of spin-polarised bulk bands in an inversion symmetric host.} Energy distribution curves from spin-resolved ARPES measurements ($h\nu=25$~eV) at the $\overline{K}$ point measured by the (a) out-of-plane ($I_\perp$) and (b) in-plane ($I_\parallel$) detectors in the Mott scattering chamber. (c) The corresponding extracted polarisations show a strong out-of-plane spin polarisation, opposite for the two valence band peaks. (d-f) The signs of all components are reversed at the $\overline{K'}$ point. (g) Our projection of the calculated bulk band structure onto the first layer of the unit cell reveals a strong spin polarisation of electronic bands localised on this layer (e.g. at $\overline{K}$), whose sign is reversed in the second layer of the unit cell as shown in (h). The measured spin polarisation at $\overline{K}$ exhibits a strong photon energy dependence (i), suggesting interference between outgoing photoelectrons originating from these different layers in the crystal.}
\end{center}
\end{figure*}

\begin{figure*}[t]
\begin{center}
\includegraphics[width=\textwidth]{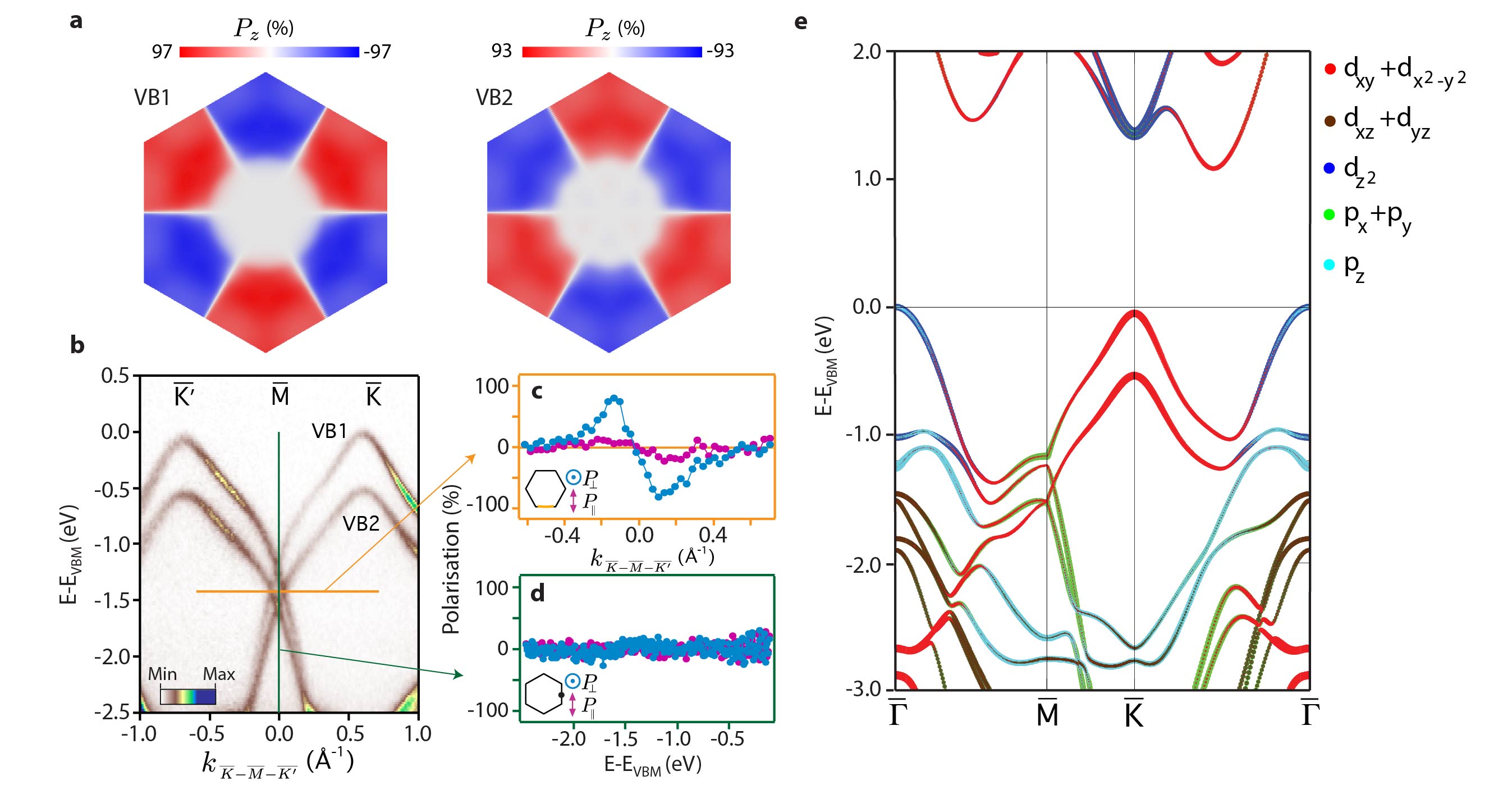}
\caption{ \label{f:KMK} {\bf Momentum-dependent suppression of layer-resolved spin polarisation.} (a) Momentum-dependence of the out-of-plane spin polarisation of the top two valence bands throughout the Brillouin zone, calculated for the $k_z=0$ plane and projected onto the first Se-W-Se layer of the unit cell. White regions indicate suppression of the layer-resolved spin-polarisation. At the $\overline{M}$ point, our (b) ARPES measurements and corresponding spin polarisation determined from spin-ARPES along the coloured (c) MDC and (d) EDC show how this occurs through the crossing of strongly spin-polarised bands. In contrast, towards the zone centre, our orbitally-projected band structure calculations (e) reveal how this is correlated with the emergence of significant out-of-plane orbital character of the electronic states.}
\end{center}
\end{figure*}

We first summarise the bulk electronic structure of WSe$_2$ (Fig.~\ref{f:overview}). The material is known to be a  semiconductor, consistent with our experimental observations where we find the Fermi level located within the band gap. We find the band extrema of the valence bands at $\Gamma$ and $K$ to be almost degenerate,~\cite{finteis_occupied_1997} but here can resolve that the valence band maximum is located at the bulk $\Gamma$ point, with significant dispersion of these zone-centre states along the surface normal ($k_z$) direction (Fig.~\ref{f:overview}(c,e)). Our measured band dispersions are in excellent agreement with those calculated from density-functional theory (DFT) (see also Supplementary Fig.~S1), confirming that we are probing the bulk electronic states of WSe$_2$. The broad total bandwidth of more than 4~eV of the cosine-like upper valence bands along $\Gamma-A$ reflects the spatially extended nature of W~$5d$ and Se~$5p$ orbitals from which these states predominantly derive. As well as these dispersive states we find a series of quasi-2D states, predominantly of planar $d_{x^2-y^2}$, $d_{xy}$, and $p_{x/y}$ orbital character (see Fig.~\ref{f:KMK}(e)). The small overlap of these orbitals along the $z$ direction results in minimal dispersion along $k_z$, while their extended nature in-plane ensures significant dispersion throughout the surface Brillouin zone (Fig.~\ref{f:overview}(d)). The lowest binding energy 2D states form a pair of hole-like bands centred at the Brillouin zone corners, contributing concentric almost circular pockets near the band top. These become trigonally warped as they grow in size with increasing binding energy, eventually merging with the zone-centre bands to form bone-shaped pockets centred at $\overline{M}$. The large splitting of $\sim\!0.5$~eV of the top of these bands at $\overline{K}$ signifies the strong atomic spin-orbit interaction in this compound, which is further reflected by our observation of hybridisation gaps, for example between two- and three-dimensional states along the $\Gamma-A$ line (Fig.~\ref{f:overview}(e)). Despite such strong spin-orbit coupling, we stress that all states remain spin-degenerate in our calculations, as expected from the bulk inversion symmetry of the crystal.

Intriguingly, however, our spin-resolved photoemission measurements reveal a strong spin polarisation of the upper pair of valence band states at the $\overline{K}$ point of the Brillouin zone (Fig.~\ref{f:Kpoint}). The measured polarisation is entirely out of the surface plane within experimental error, with up (down) orientation for the upper (lower) valence band, respectively. From fitting the measured energy distribution curves (EDCs, see methods), we estimate the magnitude of the spin polarisation to exceed 90\%, suggestive of an almost fully spin-polarised band. Moreover, the sign of all polarisations are reversed at the $\overline{K'}=-\overline{K}$ point, confirming that time-reversal symmetry remains unbroken. This points to a non-magnetic origin of the observed spin polarisation, seemingly at odds with the centrosymmetric nature of the bulk crystal structure (Fig.~\ref{f:overview}(a)). 

We attribute this to the local inversion asymmetry of each WSe$_2$ layer, leading to spin-polarised states whose texture is strongly modulated in both real and momentum space, despite the global inversion symmetry of the unit cell. For the quasi-two-dimensional bands around $\overline{K}$, our calculations reveal that the electronic wavefunctions are almost completely localised on individual Se-W-Se layers of the crystal. Within one such layer (half of the unit cell) the $D_{6h}$ symmetry of the bulk crystal is reduced to a $D_{3h}$ symmetry, allowing a net dipole moment within the $ab$-plane (Fig.~\ref{f:overview}(a)). After completion of our study, we became aware of a very recent theory that has established the general grounds by which such a lack of inversion symmetry of the crystal site point group can lead to a macroscopic spin polarisation, driven by the local nature of spin-orbit coupling.~\cite{zhang_hidden_2014} Our layer-projected calculations reveal how this causes spatial separation of spin-up and spin-down electronic states in WSe$_2$ for propagation along the $\overline{K'}-\overline{\Gamma}-\overline{K}$ direction. At $\overline{K}$, the {\it{bulk}} wavefunctions projected onto either WSe$_2$ layer of the unit cell are almost fully spin polarised for the topmost two valence bands (Fig.~\ref{f:Kpoint}(g,h)). The sign of the spin polarisation is, however, opposite between adjacent layers, thereby leading to an overall spin degeneracy of the total bulk electronic structure. 

\begin{figure*}[t]
\begin{center}
\includegraphics[width=\textwidth]{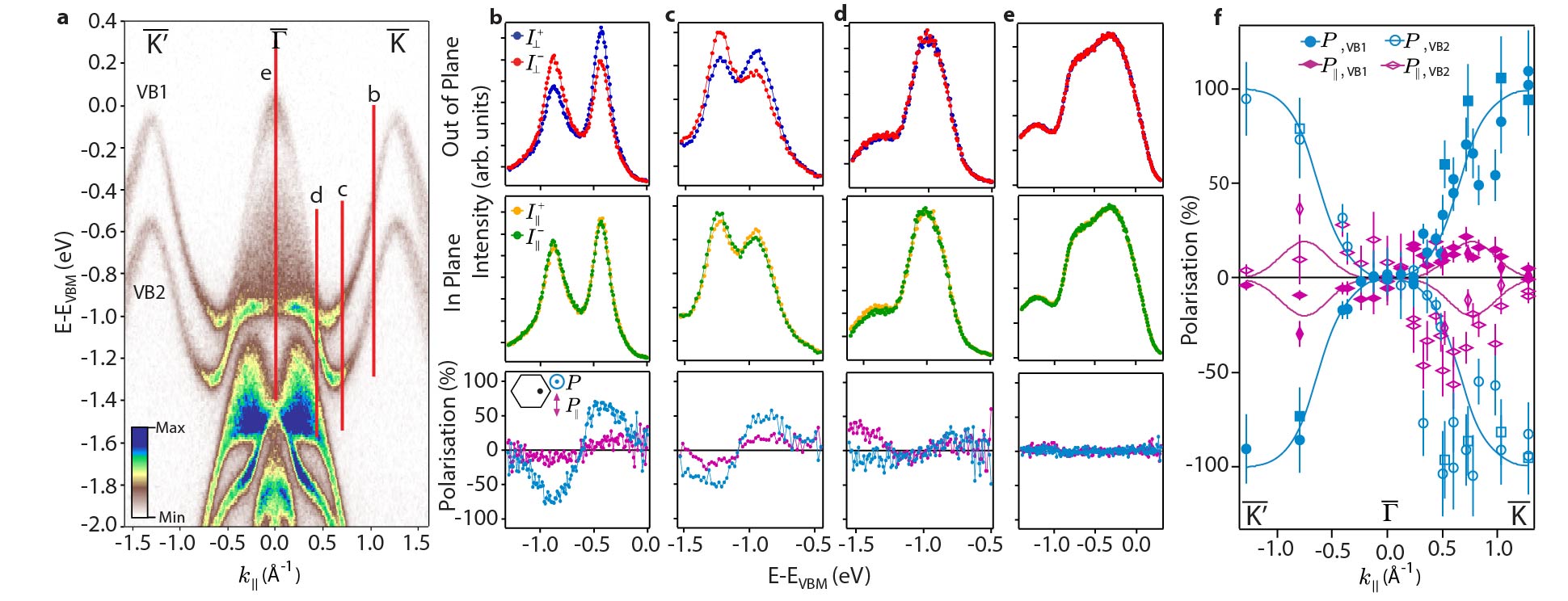}
\caption{ \label{f:GKline} {\bf Evolution of spin texture along $\overline{K'}-\overline{\Gamma}-\overline{K}$.} (a) Dispersion measured by ARPES ($h\nu=125$~eV), along the $\overline{K'}-\overline{\Gamma}-\overline{K}$ direction. Vertical lines mark the locations of the EDCs measured using spin-resolved photoemission ($h\nu=25$~eV) and corresponding extracted spin polarisations, shown in (b)--(e). The out-of- ($P_\perp$) and in- ($P_\parallel$) plane spin polarisations determined from fitting these and additional EDCs are shown in (f), revealing some canting of the spin into the surface plane away from the $\overline{K}$ points, and a total suppression of the measured spin polarisation around the zone centre. The lines in (f) are provided as guides to the eye.}
\end{center}
\end{figure*}

Photoemission, being extremely surface sensitive, can be expected to predominantly probe the top layer of this material, rationalising the strong measured spin polarisation of the bulk electronic states which we observe here. Nonetheless by changing the photon energy (Fig.~\ref{f:Kpoint}(i)), we can tune the measured polarisation nearly to zero, suggestive of the interference of spin-up and spin-down polarised photoelectrons emitted from different layers of the material,~\cite{zhu_layer-by-layer_2013,zhu_photoelectron_2014} further supporting our conclusions of a pronounced layer-dependent spin polarisation. Our calculations and experiment therefore together show how local inversion symmetry breaking leads to a huge momentum dependent spin splitting of up to $\sim\!0.5$~eV for bulk states localised in a constituent layer of the unit cell. This dramatically exceeds the typical size of spin-orbit mediated spin splittings observed to date, even in surface Rashba systems where there are strong local in-plane field gradients.~\cite{ast_giant_2007}

Figure~\ref{f:GKline} reveals how the underlying spin-polarised states evolve away from the zone corner, moving along the $\overline{K}-\overline{\Gamma}-\overline{K'}$ direction. We find a dramatic suppression of the out-of-plane spin polarisation approximately half way along this line, with negligible polarisation observed around the zone centre. This is reproduced by our {\it ab-initio} calculations (Fig.~\ref{f:KMK}(a)), and can be understood considering the orbital character of the underlying states (Fig.~\ref{f:KMK}(e)). Close to $\overline{K}$, the electronic states are derived from mostly $d_{xy}$ and $d_{x^2-y^2}$ orbitals. There is thus significant orbital overlap within the surface plane which, together with the net in-plane dipole, favours strong out-of-plane spin polarisation.~\cite{yuan_zeeman-type_2013} Around $\overline{\Gamma}$, however, the orbital character becomes dominantly $d_{z^2}$/$p_z$-like, causing this component to be strongly suppressed, as found experimentally, while also driving the observed increase in dimensionality of the electronic states. Intriguingly, we also find a small in-plane spin component emerges along $\overline{K'}-\overline{\Gamma}-\overline{K}$, which again switches sign either side of $\overline{\Gamma}$. This component would not naively be expected given the symmetry of the Se-W-Se layer which prevents dipolar fields along the out-of-plane direction. It likely reflects additional complexity beyond that considered in our theoretical approach, such as small surface relaxations leading to a non-negligible contribution of the dipole out of the surface plane. We stress, however, that this has only a small effect and the predominant contribution to the strong out-of-plane spin polarisations observed here are intrinsic to the bulk electric structure. 

We additionally find a suppression of this out-of-plane spin polarisation along the entire $\overline{M}-\overline{\Gamma}$ direction. Unlike at the zone centre, however, this cannot be attributed to a change in orbital character: The electronic states close to $\overline{M}$ are predominantly derived from planar orbitals, similar to around $\overline{K}$, and we accordingly find strong layer-resolved spin polarisations of the underlying bands close to $\overline{M}$ (Fig.~\ref{f:KMK}). Rather, the suppression of spin polarisation along $\overline{M}-\overline{\Gamma}$ is mediated by the degeneracy of two oppositely polarised bands within a single layer. At the $\overline{M}$ point itself, this is a natural consequence of time-reversal symmetry, as $\overline{M}$ is a time-reversal invariant momentum. Along the $\overline{M}-\overline{\Gamma}$ line, such degeneracies are enforced by the combination of time-reversal with the rotational $D_{3h}$ symmetry of a single monolayer within the unit cell, ensuring that the out-of-plane component of the spin must have opposite sign in neighbouring sextants of the Brillouin zone. 

Together, our calculations and experiment thus point to an extremely rich real- and momentum-space dependent spin texture of transition-metal dichalcogenides. They provide the first direct measurement of a putative spin-valley locking in these compounds, of key importance to widespread proposals to utilise these materials in exotic devices exploiting the valley pseudospin,\cite{xiao_coupled_2012,mak_control_2012,zeng_valley_2012,jones_optical_2013,zhang_electrically_2014} and open new opportunities to manipulate and probe the layer pseudospin.\cite{layer_p} They also reveal additional in-plane spin canting away from the band extrema, which may explain the loss in the degree of valley polarisation that can be induced with increasing energy of photoexcitation.~\cite{mak_control_2012,jones_optical_2013} More generally, our experimental realisation of a system where bulk inversion symmetry breaking is not a necessary condition to stabilise spin-polarised electronic states opens a wealth of new opportunities for creating and controlling spin and valley polarisation in bulk solids via local inversion asymmetry.\\

{\small 
\noindent{\bf Methods}\\
\noindent{\it ARPES:} ARPES measurements were performed at the I05 beamline of Diamond Light Source, UK, and spin-ARPES measurements at the I3 beamline of MAX-III synchrotron, Sweden.~\cite{berntsen_spin-_2010}  Single crystal samples of WSe$_2$, grown by the chemical vapour transport method, were cleaved {\it in-situ} and measured at temperatures ranging from 30-300~K. Measurements were performed using $p$-polarised synchrotron light from 20-130~eV (ARPES) and 20-40~eV (spin-ARPES), and employing Scienta R4000 hemispherical electron analysers. For the spin-ARPES measurements, a mini-Mott detector scheme was utilised, permitting simultaneous detection of the out-of-plane and one in-plane (along the analyser slit direction) component of the photoelectron spin.~\cite{berntsen_spin-_2010} A Sherman function of $S=0.17$ was used to generate the measured spin polarisations,~\cite{berntsen_spin-_2010} 
\[P_i=\frac{(I_i^+-I_i^-)}{S(I_i^++I_i^-)},\]
where $P_i$ is the photoelectron spin polarisation measured along the out-of-plane, $i=\perp$, or in-plane, $i=\parallel$, direction, and $I_i^\pm$ is the measured intensity on the individual detectors in the Mott scattering chamber, corrected by a relative detector efficiency calibration. To extract numerical values of the polarisation, we fitted the measured EDCs to two Lorentzian peaks and a Shirley background, convolved with a Gaussian function to account for the instrumental resolution, with the corresponding Lorentzian peak areas used to derive the measured spin polarisation. We also applied a geometrical correction to account for the finite angle between the sample and the electron spectrometer, and the corresponding influence of this on the spin polarisation measured in the reference frame of the spectrometer. To determine the $k_z$ dispersion from photon-energy dependent ARPES, we employed a free electron final state model 
\[k_z=\sqrt{2m_e / \hbar^2}(V_0 + E_{\mathrm{k}} \cos^2 \theta)^{1/2},\] 
where $\theta$ is the in-plane emission angle and $V_0$ is the inner potential. Our photon energy range covers more than 6 complete Brillouin zones along $k_z$, and we find best agreement taking an inner potential of 18~eV.

\noindent{\it Calculations:} Electronic structure calculations were performed within the context of density functional theory (DFT) using the modified Becke-Johnson exchange potential and Perdew-Burke-Ernzerhof correlation functional as implemented in the WIEN2K programme.~\cite{wien2k} Relativistic effects, including spin-orbit coupling, were fully included. The Brillouin zone was sampled by a 12x12x6 $k$-mesh. For the orbital and layer projection calculation, a  tight binding Hamiltonian for the bulk band structure was constructed by downfolding the DFT results using maximally localised Wannier functions,~\cite{souza,mostofi,kunes} employing as a basis W $5d$ and $5s$ orbitals and Se $5p$ and $5s$ orbitals.}\\
\

\small{
\noindent{\bf Acknowledgements:} We gratefully acknowledge support from the Engineering and Physical Sciences Research Council, UK, the VILLUM foundation, the Calipso program, TRF-SUT Grant RSA5680052 and  NANOTEC, Thailand through the CoE Network. PDCK acknowledges support from the Royal Society through a University Research Fellowship. MSB was supported by the Grant-in-Aid for Scientific Research (S) (No. 24224009) from the Ministry of Education, Culture, Sports, Science and Technology (MEXT) of Japan. The experiments at the MAX IV Laboratory were made possible through funding from the Swedish Research Council and the Knut and Alice Wallenberg Foundation. }
\newpage
\bibliographystyle{naturemag}

\begin{thebibliography}{30}
\expandafter\ifx\csname url\endcsname\relax
  \def\url#1{\texttt{#1}}\fi
\expandafter\ifx\csname urlprefix\endcsname\relax\def\urlprefix{URL }\fi
\providecommand{\bibinfo}[2]{#2}
\providecommand{\eprint}[2][]{\url{#2}}

\bibitem{koo_control_2009}
\bibinfo{author}{Koo, H.~C.} \emph{et~al.}
\newblock \bibinfo{title}{Control of Spin Precession in a Spin-Injected Field
  Effect Transistor}.
\newblock \emph{\bibinfo{journal}{Science}} \textbf{\bibinfo{volume}{325}},
  \bibinfo{pages}{1515-1518} (\bibinfo{year}{2009}).
%\newblock \bibinfo{note}{Bibtex: Koo2009}.

\bibitem{hsieh_topological_2008}
\bibinfo{author}{Hsieh, D.} \emph{et~al.}
\newblock \bibinfo{title}{A Topological Dirac Insulator in a Quantum Spin Hall
  Phase}.
\newblock \emph{\bibinfo{journal}{Nature}} \textbf{\bibinfo{volume}{452}},
  \bibinfo{pages}{970-974} (\bibinfo{year}{2008}).
%\newblock \bibinfo{note}{Bibtex: Hsieh2008}.

\bibitem{Bahramy:NatCommun:3(2012)1159--}
\bibinfo{author}{Bahramy, M.} \emph{et~al.}
\newblock \bibinfo{title}{Emergent Quantum Confinement at Topological Insulator
  Surfaces}.
\newblock \emph{\bibinfo{journal}{Nature Commun.}}
  \textbf{\bibinfo{volume}{3}}, \bibinfo{pages}{1159} (\bibinfo{year}{2012}).

\bibitem{Murakawa_Detection_2013}
\bibinfo{author}{Murakawa, H.} \emph{et~al.}
\newblock \bibinfo{title}{Detection of Berry's Phase in a Bulk Rashba Semiconductor}.
\newblock \emph{\bibinfo{journal}{Science}}
  \textbf{\bibinfo{volume}{342}}, \bibinfo{pages}{1490--1493} (\bibinfo{year}{2013}).
  
  \bibitem{mourik_signatures_2012}
\bibinfo{author}{Mourik, V.} \emph{et~al.}
\newblock \bibinfo{title}{Signatures of Majorana Fermions in Hybrid
  Superconductor-Semiconductor Nanowire Devices}.
\newblock \emph{\bibinfo{journal}{Science}} \textbf{\bibinfo{volume}{336}},
  \bibinfo{pages}{1003--1007} (\bibinfo{year}{2012}).
%\newblock \urlprefix\url{http://www.sciencemag.org/content/336/6084/1003}.
%\newblock \bibinfo{note}{{PMID:} 22499805}.

  \bibitem{zhang_hidden_2014}
\bibinfo{author}{Zhang, X.}, \bibinfo{author}{Liu, Q.}, \bibinfo{author}{Luo,
  J.~W.}, \bibinfo{author}{Freeman, A.~J.} \& \bibinfo{author}{Zunger, A.}
\newblock \bibinfo{title}{Hidden Spin Polarization in Inversion-Symmetric Bulk
  Crystals}.
\newblock \emph{\bibinfo{journal}{Nature Phys.}}
  \textbf{\bibinfo{volume}{10}}, \bibinfo{pages}{387--393}
  (\bibinfo{year}{2014}).
%\newblock
  %\urlprefix\url{http://www.nature.com/nphys/journal/v10/n5/full/nphys2933.html}.

\bibitem{mak_control_2012}
\bibinfo{author}{Mak, K.~F.}, \bibinfo{author}{He, K.}, \bibinfo{author}{Shan,
  J.} \& \bibinfo{author}{Heinz, T.~F.}
\newblock \bibinfo{title}{Control of Valley Polarization in Monolayer MoS$_2$ by Optical Helicity}.
\newblock \emph{\bibinfo{journal}{Nature Nano.}}
  \textbf{\bibinfo{volume}{7}}, \bibinfo{pages}{494--498}
  (\bibinfo{year}{2012}).
%\newblock
  %\urlprefix\url{http://www.nature.com/nnano/journal/v7/n8/full/nnano.2012.96.html}.

\bibitem{zeng_valley_2012}
\bibinfo{author}{Zeng, H.}, \bibinfo{author}{Dai, J.}, \bibinfo{author}{Yao,
  W.}, \bibinfo{author}{Xiao, D.} \& \bibinfo{author}{Cui, X.}
\newblock \bibinfo{title}{Valley Polarization in MoS$_2$ Monolayers by Optical Pumping}.
\newblock \emph{\bibinfo{journal}{Nature Nano.}}
  \textbf{\bibinfo{volume}{7}}, \bibinfo{pages}{490--493}
  (\bibinfo{year}{2012}).
%\newblock
  %\urlprefix\url{http://www.nature.com/nnano/journal/v7/n8/full/nnano.2012.95.html}.


\bibitem{zhang_electrically_2014}
\bibinfo{author}{Zhang, Y.~J.}, \bibinfo{author}{Oka, T.},
  \bibinfo{author}{Suzuki, R.}, \bibinfo{author}{Ye, J.~T.} \&
  \bibinfo{author}{Iwasa, Y.}
\newblock \bibinfo{title}{Electrically Switchable Chiral Light-Emitting
  Transistor}.
\newblock \emph{\bibinfo{journal}{Science}} \textbf{\bibinfo{volume}{344}},
  \bibinfo{pages}{725--728} (\bibinfo{year}{2014}).
%\newblock \urlprefix\url{http://www.sciencemag.org/content/344/6185/725}.
%\newblock \bibinfo{note}{{PMID:} 24790028}.

\bibitem{datta_electronic_1990}
\bibinfo{author}{Datta, S.} \& \bibinfo{author}{Das, B.}
\newblock \bibinfo{title}{Electronic Analog of the Electro-Optic Modulator}.
\newblock \emph{\bibinfo{journal}{Appl. Phys. Lett.}}
  \textbf{\bibinfo{volume}{56}}, \bibinfo{pages}{665-667}
  (\bibinfo{year}{1990}).
%\newblock \bibinfo{note}{Bibtex: Datta1990}.

\bibitem{das_engineering_2013}
\bibinfo{author}{Das, T.} \& \bibinfo{author}{Balatsky, A.~V.}
\newblock \bibinfo{title}{Engineering Three-Dimensional Topological Insulators
  in Rashba-Type Spin-Orbit Coupled Heterostructures}.
\newblock \emph{\bibinfo{journal}{Nature Commun.}}
  \textbf{\bibinfo{volume}{4}}, \bibinfo{pages}{1972}
   (\bibinfo{year}{2013}).
%\newblock
  %\urlprefix\url{http://www.nature.com/ncomms/2013/130606/ncomms2972/full/ncomms2972.html}.

\bibitem{zhou_engineering_2014}
\bibinfo{author}{Zhou, J.~J.}, \bibinfo{author}{Feng, W.},
  \bibinfo{author}{Zhang, Y.}, \bibinfo{author}{Yang, S.~A.} \&
  \bibinfo{author}{Yao, Y.}
\newblock \bibinfo{title}{Engineering Topological Surface States and Giant
  Rashba Spin Splitting in BiTeI/Bi$_2$Te$_3$ Heterostructures}.
\newblock \emph{\bibinfo{journal}{Sci. Rep.}}
  \textbf{\bibinfo{volume}{4}}, \bibinfo{pages}{3841}
   (\bibinfo{year}{2014}).
%\newblock
%  \urlprefix\url{http://www.nature.com/srep/2014/140123/srep03841/full/srep03841.html}.

\bibitem{bychkov_properties_1984}
\bibinfo{author}{Bychkov, Y.~A.} \& \bibinfo{author}{Rashba, E.~I.}
\newblock \bibinfo{title}{Properties of a 2D Electron Gas with Lifted
  Spectral Degeneracy}.
\newblock \emph{\bibinfo{journal}{{JETP} Lett.}} \textbf{\bibinfo{volume}{39}},
  \bibinfo{pages}{78-81} (\bibinfo{year}{1984}).
%\newblock \bibinfo{note}{Bibtex: Bychkov1984}.


\bibitem{nitta_gate_1997}
\bibinfo{author}{Nitta, J.}, \bibinfo{author}{Akazaki, T.},
  \bibinfo{author}{Takayanagi, H.} \& \bibinfo{author}{Enoki, T.}
\newblock \bibinfo{title}{Gate Control of Spin-Orbit Interaction in an Inverted
  In$_{\textrm{0.53}}$Ga$_{\textrm{0.47}}$As/In$_{\textrm{0.52}}$Al$_{\textrm{0.48}}$As
  Heterostructure}.
\newblock \emph{\bibinfo{journal}{Phys. Rev. Lett.}}
  \textbf{\bibinfo{volume}{78}}, \bibinfo{pages}{1335--1338}
  (\bibinfo{year}{1997}).
%\newblock \bibinfo{note}{Bibtex: Nitta1997}.

\bibitem{ast_giant_2007}
\bibinfo{author}{Ast, C.~R.} \emph{et~al.}
\newblock \bibinfo{title}{Giant Spin Splitting through Surface Alloying}.
\newblock \emph{\bibinfo{journal}{Phys. Rev. Lett.}}
  \textbf{\bibinfo{volume}{98}}, \bibinfo{pages}{186807}
  (\bibinfo{year}{2007}).
%\newblock \bibinfo{note}{Bibtex: Zhu2013}.

\bibitem{king_large_2011}
\bibinfo{author}{King, P.~D.~C.} \emph{et~al.}
\newblock \bibinfo{title}{Large Tunable Rashba Spin Splitting of a
  Two-Dimensional Electron Gas in Bi$_2$Se$_3$}.
\newblock \emph{\bibinfo{journal}{Phys. Rev. Lett.}}
  \textbf{\bibinfo{volume}{107}}, \bibinfo{pages}{096802}
  (\bibinfo{year}{2011}).
%\newblock \bibinfo{note}{Bibtex: King2011-2}.

\bibitem{king_STO}
\bibinfo{author}{King, P. D.~C.} \emph{et~al.}
\newblock \bibinfo{title}{Quasiparticle Dynamics and Spin-Orbital Texture of the SrTiO$_3$ Two-Dimensional Electron Gas}.
\newblock \emph{\bibinfo{journal}{Nature Commun.}}
  \textbf{\bibinfo{volume}{5}}, \bibinfo{pages}{3414}
  (\bibinfo{year}{2014}).
  
\bibitem{dresselhaus_spin-orbit_1955}
\bibinfo{author}{Dresselhaus, G.}
\newblock \bibinfo{title}{Spin-Orbit Coupling Effects in Zinc Blende
  Structures}.
\newblock \emph{\bibinfo{journal}{Phys. Rev.}}
  \textbf{\bibinfo{volume}{100}}, \bibinfo{pages}{580-586}
  (\bibinfo{year}{1955}).
%\newblock \urlprefix\url{http://link.aps.org/doi/10.1103/PhysRev.100.580}.

\bibitem{ishizaka_giant_2011}
\bibinfo{author}{Ishizaka, K.} \emph{et~al.}
\newblock \bibinfo{title}{Giant Rashba-Type Spin Splitting in Bulk BiTeI}.
\newblock \emph{\bibinfo{journal}{Nature Mat.}}
  \textbf{\bibinfo{volume}{10}}, \bibinfo{pages}{521--526}
  (\bibinfo{year}{2011}).
%\newblock
%  \urlprefix\url{http://www.nature.com/nmat/journal/v10/n7/full/nmat3051.html}.


\bibitem{xiao_coupled_2012}
\bibinfo{author}{Xiao, D.}, \bibinfo{author}{Liu, G.-B.},
  \bibinfo{author}{Feng, W.}, \bibinfo{author}{Xu, X.} \& \bibinfo{author}{Yao,
  W.}
\newblock \bibinfo{title}{Coupled Spin and Valley Physics in Monolayers of
  MoS$_2$ and Other Group-VI Dichalcogenides}.
\newblock \emph{\bibinfo{journal}{Phys. Rev. Lett.}}
  \textbf{\bibinfo{volume}{108}}, \bibinfo{pages}{196802}
  (\bibinfo{year}{2012}).
%\newblock
  %\urlprefix\url{http://link.aps.org/doi/10.1103/PhysRevLett.108.196802}.


\bibitem{finteis_occupied_1997}
\bibinfo{author}{Finteis, T.} \emph{et~al.}
\newblock \bibinfo{title}{Occupied and Unoccupied Electronic Band Structure of
  WSe$_2$}.
\newblock \emph{\bibinfo{journal}{Phys. Rev. B}}
  \textbf{\bibinfo{volume}{55}}, \bibinfo{pages}{10400--10411}
  (\bibinfo{year}{1997}).
%\newblock \urlprefix\url{http://link.aps.org/doi/10.1103/PhysRevB.55.10400}.


\bibitem{zhu_layer-by-layer_2013}
\bibinfo{author}{Zhu, Z.-H.} \emph{et~al.}
\newblock \bibinfo{title}{Layer-by-Layer Entangled Spin-Orbital Texture of the
  Topological Surface State in Bi$_2$Se$_3$}.
\newblock \emph{\bibinfo{journal}{Phys. Rev. Lett.}}
  \textbf{\bibinfo{volume}{110}}, \bibinfo{pages}{216401}
  (\bibinfo{year}{2013}).
%\newblock \bibinfo{note}{Bibtex: Zhu2013}.

\bibitem{zhu_photoelectron_2014}
\bibinfo{author}{Zhu, Z.-H.} \emph{et~al.}
\newblock \bibinfo{title}{Photoelectron Spin-Polarization Control in the
  Topological Insulator Bi$_2$Se$_3$}.
\newblock \emph{\bibinfo{journal}{Phys. Rev. Lett.}}
  \textbf{\bibinfo{volume}{112}}, \bibinfo{pages}{076802}
  (\bibinfo{year}{2014}).
%\newblock
%  \urlprefix\url{http://link.aps.org/doi/10.1103/PhysRevLett.112.076802}.



\bibitem{yuan_zeeman-type_2013}
\bibinfo{author}{Yuan, H.} \emph{et~al.}
\newblock \bibinfo{title}{Zeeman-Type Spin Splitting Controlled by an Electric
  Field}.
\newblock \emph{\bibinfo{journal}{Nature Phys.}}
  \textbf{\bibinfo{volume}{9}}, \bibinfo{pages}{563--569}
  (\bibinfo{year}{2013}).
%\newblock
%  \urlprefix\url{http://www.nature.com/nphys/journal/v9/n9/full/nphys2691.html}.
%


\bibitem{jones_optical_2013}
\bibinfo{author}{Jones, A.~M.} \emph{et~al.}
\newblock \bibinfo{title}{Optical Generation of Excitonic Valley Coherence in
  Monolayer WSe$_2$}.
\newblock \emph{\bibinfo{journal}{Nature Nano.}}
  \textbf{\bibinfo{volume}{8}}, \bibinfo{pages}{634--638}
  (\bibinfo{year}{2013}).
%\newblock
  %\urlprefix\url{http://www.nature.com/nnano/journal/v8/n9/full/nnano.2013.151.html}.


\bibitem{layer_p}
\bibinfo{author}{Xu, X.}, \bibinfo{author}{Yao, W.}, \bibinfo{author}{Xiao, D.} \& \bibinfo{author}{Heinz, T.~F.}
\newblock \bibinfo{title}{Spin and Pseudospins in Layered Transition Metal Dichalcogenides}.
\newblock \emph{\bibinfo{journal}{Nature}}
  \textbf{\bibinfo{volume}{10}}, \bibinfo{pages}{343-350}
  (\bibinfo{year}{2014}).

\bibitem{berntsen_spin-_2010}
\bibinfo{author}{Berntsen, M.~H.} \emph{et~al.}
\newblock \bibinfo{title}{A Spin- and Angle-Resolving Photoelectron
  Spectrometer}.
\newblock \emph{\bibinfo{journal}{Rev. Sci. Instrum.}}
  \textbf{\bibinfo{volume}{81}}, \bibinfo{pages}{035104}
  (\bibinfo{year}{2010}).
%\newblock
 % \urlprefix\url{http://scitation.aip.org/content/aip/journal/rsi/81/3/10.1063/1.3342120}.

 \bibitem{wien2k} 
Blaha, P. {et al.}, WIEN2K package, Version 10.1 (2010); available at, http://www.wien2k.at.

\bibitem{souza}
\bibinfo{author}{Souza, I.} \emph{et~al.}
\newblock \bibinfo{title}{Maximally Localized Wannier Functions for Entangled Energy Bands}.
\newblock \emph{\bibinfo{journal}{Phys. Rev. B}} \textbf{\bibinfo{volume}{65}},
  \bibinfo{pages}{035109} (\bibinfo{year}{2001}).  

\bibitem{mostofi} 
\bibinfo{author}{Mostofi, A. A.} \emph{et~al.} 
\newblock \bibinfo{title}{Wannier90: A Tool for Obtaining Maximally Localised Wannier Functions}.
\newblock \emph{\bibinfo{journal}{Comp. Phys. Commun.}}
\textbf{\bibinfo{volume}{178}}, 
\bibinfo{pages}{685Ð699} (\bibinfo{year}{2008}).
 
   \bibitem{kunes}
\bibinfo{author}{Kune\v s, J.} \emph{et~al.}
\newblock \bibinfo{title}{WIEN2WANNIER: From Linearized Augmented Plane Waves to Maximally Localized Wannier Functions}.
\newblock \emph{\bibinfo{journal}{Comp. Phys. Commun. }} \textbf{\bibinfo{volume}{181}},
  \bibinfo{pages}{1888Ð1895} (\bibinfo{year}{2010}). 


\end{thebibliography}

\end{document}